\documentclass[hyper,letterpaper,12pt]{article}
\usepackage{a4wide}
\usepackage{amsmath}
\usepackage[
      colorlinks=true,
      linkcolor=blue,
      urlcolor=blue,
      filecolor=black,
      citecolor=blue,
            ]{hyperref}
\usepackage{color}
\usepackage{graphicx}
\usepackage{amsfonts}
\usepackage{amssymb}
\usepackage{epstopdf}

\def\beq{\begin{equation}}
\def\eeq{\end{equation}}

\begin{document}
\title{\bf \Large A Holographic Study on Vector Condensate Induced by a Magnetic Field}
%Magnetic Field Induced Instability In a Holographic Model With a Charged Vector Condensate}

\author{\large
~Rong-Gen Cai$^1$\footnote{E-mail: cairg@itp.ac.cn}~,
~~Song He$^1$\footnote{E-mail: hesong@itp.ac.cn}~,
~~Li Li$^1$\footnote{E-mail: liliphy@itp.ac.cn}~,
~~Li-Fang Li$^2$\footnote{E-mail: lilf@itp.ac.cn}\\
\\
\small $^1$State Key Laboratory of Theoretical Physics,\\
\small Institute of Theoretical Physics, Chinese Academy of Sciences,\\
\small Beijing 100190,  China.\\
\small $^2$State Key Laboratory of Space Weather, \\
\small Center for Space Science and Applied Research, Chinese Academy of Sciences,\\
\small Beijing 100190, China.}
\date{\today}
\maketitle

\begin{abstract}
\normalsize We study a holographic model with vector condensate by coupling the anti-de Sitter gravity to an Abelian gauge field and a charged vector field in $(3+1)$ dimensional spacetime. In this model there exists a non-minimal coupling of the vector field to the gauge field.  We find that there is a critical temperature below which the charged vector condenses via a second order phase transition. The DC conductivity becomes infinite and the AC conductivity develops a gap in the condensed phase. We study the effect of a background magnetic field on the system. It is found that the background magnetic field can induce the condensate of the vector field even in the case without chemical potential/charge density. In the case with non-vanishing charge density, the transition temperature raises with the applied magnetic field, and the condensate of the charged vector operator forms a vortex lattice structure in the spatial directions perpendicular to the magnetic field.
\end{abstract}

\tableofcontents

\section{ Introduction}
The gauge/gravity duality~\cite{Maldacena:1997re,Gubser:1998bc,Witten:1998qj} turns out to be a useful and complimentary framework to study a strongly coupled system through an appropriate gravity theory living in a higher dimensional spacetime. There are two complementary approaches. In the top-down approach, one can obtain some low energy effective theories in the bulk as consistent truncations of string/M theory by dimensional reduction on some compactified manifolds. The advantage of this approach is that one knows the origin of the effective gravity theory in the bulk and the details of the dual field theory. One of the well-known examples in this category is the duality between the IIB superstring theory on the $AdS_5 \times S^5$ and the ${\cal N}=4$ supersymmetric Yang-Mills theory on the $AdS_5$ boundary. On the other hand, in the so-called bottom-up approach, the gravity theory in the hulk is usually constructed by some physical considerations, according to the AdS/CFT dictionary. The models constructed in this way are simple and universal, just like the Landau-Ginzburg theory describing superconductivity. The disadvantage is that the details of the dual field theory are not very clear.

A well known example in the second category is the holographic superconductor model~\cite{Hartnoll:2008vx,Hartnoll:2008kx}. More specifically, to build a holographic
superconductor model, one should first introduce a U(1) gauge field in the bulk, which corresponds to a global U(1) symmetry in the boundary side. To break this U(1) symmetry spontaneously, one needs a charged operator condensing at low temperature. Therefore, one includes a charged scalar field in the bulk which is dual to the boundary scalar operator. For simplicity, the gauge field and the charged scalar can be minimally coupled. This holographic toy model admits black holes with scalar hair at low temperatures (superconducting phase), but without scalar hair at high temperatures (normal phase). In this way, one mimics the superconductor/conductor phase transition in the field theory side.  Holographic superconductor models constructed in the top-down approach can also be found, for example, in refs.~\cite{Gubser:2009qm,Gauntlett:2009dn,Ammon:2008fc,Bobev:2011rv}.

It is well known that superconductivity involves the formation of a quantum condensate state by pairing conduction electrons.  The pair of electrons, called the Cooper pair, can be in a
state of either spin singlet with total spin $s=0$ or spin triplet with $s=1$. The order parameter in a superconductor is expressed in terms of the gap function. Due to the anti-commuting properties of the electron wave function, the antisymmetric spin-singlet state is associated with a symmetric orbital wave function with orbital angular momentum $l=0$ (s-wave), 2 (d-wave), etc, while the symmetric spin-triplet state is accompanied by an antisymmetric orbital wave function with orbital angular momentum $l=1$ (p-wave), 3 (f-wave), etc. Since the condensed field in the holographic setup~\cite{Hartnoll:2008vx,Hartnoll:2008kx} is a scalar field dual to a scalar operator in the field theory side, it is therefore a holographic s-wave model. In the holographic setup, a d-wave order parameter is dual to a charged  spin two field propagating in the bulk~\cite{Chen:2010mk,Benini:2010pr}. To mimic p-wave superconductor, one can consider a vector order parameter which is dual to a vector field in the gravity side. A holographic p-wave model~\cite{Gubser:2008wv} was constructed by adding a SU(2) Yang-Mills field into the bulk. A U(1) subgroup of the Yang-Mills field is regarded as the gauge group of electromagnetism. A gauge boson generated by another SU(2) generator charged under this U(1) by the nonlinear coupling of the non-Abelian field is dual to the vector order parameter. An alternative holographic realization of p-wave superconductivity emerges from the condensation of a 2-form field in the bulk~\cite{Aprile:2010ge}. A holographic chiral $p_x+ip_y$ superconductor was discussed by generalized the SU(2) model introducing a Maxwell field and a Chern-Simons term~\cite{Zayas:2011dw}. Note that in principle we can also build a holographic  p-wave model by introducing a complex vector field charged under a U(1) gauge field in the bulk with possible couplings.

On the other hand, motivated by the possibility to create a very strong magnetic field, for example, at RHIC and LHC, the interest to investigate the properties of QCD matter in a strong magnetic field has been growing recently. Some new phenomena have been revealed such as the chiral magnetic effect, a split between the deconfinement phase transition and chiral symmetry restoration phase trasnition, see ref.~\cite{Kharzeev:2012ph} for a review.  An interesting new phenomenon is the possibility that the QCD vacuum undergoes a phase transition to a new phase with charged $\rho$-meson condensed in a sufficiently strong magnetic field~\cite{Chernodub:2010qx,Chernodub:2011mc}. This exotic phase is a kind of anisotropic superconducting phase~\cite{Chernodub:2011gs}. The author in ref.~\cite{Chernodub:2010qx} adopted an effective quantum electrodynamics action (DSGS model~\cite{Djukanovic:2005ag}) to discuss the condensate of  $\rho$-meson. A similar study based on Nambu-Jona-Lasinio model~\cite{Nambu:1961tp} can be found in ref.~\cite{Chernodub:2011mc}. The uniform magnetic field background is encoded in an Abelian gauge field. As a vector boson, the possible condensate of the $\rho$-meson in a uniform magnetic field can be realized in the holographic framework by introducing a charged vector field in the gravity side in the presence of a magnetic field. Based on the Sakai-Sugimoto model, the holographic $\rho$-meson was studied in the confinement phase at zero temperature~\cite{Callebaut:2011ab} where it is shown that the effective mass of the $\rho$-meson at strong magnetic field becomes tachyonic. The influence of the magnetic field on the chiral transition temperature and deconfinement transition temperature was discussed in ref.~\cite{Callebaut:2013ria}.  In the SU(2) model~\cite{Gubser:2008wv}, a similar instability triggered by a non-Abelian magnetic field has been found in refs.~\cite{Ammon:2011je,Bu:2012mq,Wong:2013rda}, which is reminiscent of the observation that non-Abelian magnetic field induces the W-boson condensate exhibiting vortex lattices in flat spacetime~\cite{Nielsen:1978rm,Ambjorn:1988fx,Ambjorn:1988gb}.

In this paper we will construct a toy model by introducing a complex vector field charged under an Abelian gauge field in the bulk, which is dual to a strongly coupled system involving a charged vector operator with a global U(1) symmetry. In this model there exists a non-minimal coupling between the vector field and the Abelian gauge field.  In this bottom-up approach, we do not know the details of the dual field theory clearly. However, our setup meets the minimal requirement to construct a holographic p-wave superconductor model.  This model has also potential to discuss the $\rho$-meson condensate in the presence of magnetic field.

We first study the model in the case without a background magnetic field, i.e., turn on the temporal component of the gauge field only, which is the usual way of analyzing holographic superconductor at a first step~\cite{Hartnoll:2008vx,Chen:2010mk,Benini:2010pr,Gubser:2008wv}. We find a critical temperature below which a vector operator acquires a vacuum expectation value breaking the Abelian symmetry spontaneously. Furthermore, the condensate of this vector operator picks out a special spatial direction, thus the rotational symmetry is also broken in the condensed phase. The system undergoes a second order phase transition with the critical exponent one half which coincides with the result from mean field theory. This condensed phase presents characteristics known from superconductivity, such as an infinite DC conductivity and a gap in the optical conductivity.

An instability induced by a background non-Abelian magnetic field has been reported, where the non-Abelian current operators obtain vacuum expectation values resulting in a vortex lattice structure~\cite{Bu:2012mq,Wong:2013rda}. We are interested in how the applied Abelian magnetic field influences the instability of our model with Abelian gauge symmetry. For this, we turn on a uniform magnetic field $B$ in the bulk, which immerses the condensed phase into an external magnetic field.

We find that the increase of the magnetic field induces the instability of the black hole background, which gives rise to a family of condensate induced by the applied magnetic field in the dual strongly coupled system. This magnetic field induced instability can happen even for the case with vanishing chemical potential or charge density. For finite magnetic field, there is a tower of ``droplet" solutions~\cite{Albash:2008eh} in the sense that they are localized in a  finite region. Further, the condensate shrinks in size as one increases the magnetic field, which shares similarity with the Meissner effect. But it is not the exact case with conventional superconductivity, where the superconductivity is suppressed by magnetic field in the real materials. Of course, it is interesting to notice the fact that it has been reported recently that an applied magnetic field may induce superconductivity~\cite{levy2005,uji2010}.  In addition, the emergence of charged vector operator condensate triggered by the applied magnetic field is consistent with the appearance of electromagnetically superconducting phase in a strong magnetic field studied from both field theory method~\cite{Chernodub:2010qx,Chernodub:2011mc} and the holographic setup~\cite{ Callebaut:2011ab,Callebaut:2013ria}. We also manage to construct the vortex lattice solution in the condensed phase near the phase transition, which is very reminiscent of the Abrikosov lattice in common type-II superconductors.

The plan of this paper is as follows. In section~\ref{sect:model}, we introduce the holographic model and deduce the general equations of motion of the system. In section~\ref{sect:ansatzB}, we turn on a uniform magnetic field and give details about how to recover the Landau levels in this holographic model. Section~\ref{sect:diagram} is devoted to discussing the phase diagram with/without chemical potential. The vortex lattice solutions are constructed in Section~\ref{sect:vortex}. The conclusion and further discussions are included in section~\ref{sect:conclusion}. We study, in Appendix~\ref{sect:superconducor}, the condensate of the vector field and the phase transition in the case without magnetic field by calculating the conductivity to ensure  the condensed phase to be a superconducting state.
%%
%%%%%
%%%section%2%%%%%%%%%%%%%%%%%%%%%%%%
%%\section%2%P-Wave Model
%%%section%2%%%%%%%%%%%%%%%%%%%%%%%
%%%%%%%%%
%_________________________________________________________________%

\section{The holographic model}
\label{sect:model}

Let us introduce a charged vector field into the $(3+1)$ dimensional Einstein-Maxwell theory with a negative cosmological constant. The full action reads
\begin{equation}\label{action}
S =\frac{1}{2\kappa^2}\int d^4 x
\sqrt{-g}[\mathcal{R}+\frac{6}{L^2}-\frac{1}{4}F_{\mu\nu} F^{\mu \nu}-\frac{1}{2}\rho_{\mu\nu}^\dagger\rho^{\mu\nu}-m^2\rho_\mu^\dagger\rho^\mu+iq\gamma \rho_\mu\rho_\nu^\dagger F^{\mu\nu}],
\end{equation}
where $L$ is the AdS radius that we will set to be unity, $\kappa^2\equiv 8\pi G $ is related to the gravitational constant in the bulk and $m$ is the mass of the charged vector field $\rho_\mu$. The strength of U(1) field $A_\mu$ is $F_{\mu\nu}=\nabla_\mu A_\nu-\nabla_\nu A_\mu$. The tensor  $\rho_{\mu\nu}$ in~\eqref{action} is defined by $\rho_{\mu\nu}=D_\mu\rho_\nu-D_\nu\rho_\mu$. The covariant derivative $D_\mu=\nabla_\mu-iq A_\mu$ with $q$ the charge of vector $\rho_\mu$. The last interacting term describes the non-minimal coupling of the charged vector field $\rho_\mu$ to the U(1) gauge field $A_\mu$. The parameter $\gamma$ characterizes the magnetic moment of the vector field $\rho_\mu$ and is assumed to be non-negative. The form of the action is reminiscent of the DSGS model which describes the quantum electrodynamics of the $\rho$-meson proposed by Djukanovic, Schindler, Gegelia, and Scherer in ref.~\cite{Djukanovic:2005ag}. Compared to the action of the DSGS model, it is easy to find that the action in~\eqref{action} is just a simple generalization of the DSGS model to the anti-de Sitter space. Note that the part of neutral $\rho$-meson in ref.~\cite{Djukanovic:2005ag} is neglected here since it is not relevant to our goal in this paper.

Varying the action~\eqref{action} with respect to $A_\mu$ yields the equation of motion for gauge field
\begin{equation}\label{gauge}
\nabla^\nu F_{\nu\mu}=iq(\rho^\nu\rho_{\nu\mu}^\dagger-{\rho^\nu}^\dagger\rho_{\nu\mu})+iq\gamma\nabla^\nu(\rho_\nu\rho_\mu^\dagger-\rho_\nu^\dagger\rho_\mu),
\end{equation}
while a variation of the action~\eqref{action} with respect to ${\rho^\nu}^\dagger$  gives the equation of motion for the charged vector field
\begin{equation}\label{vector}
D^\nu\rho_{\nu\mu}-m^2\rho_\mu+iq\gamma\rho^\nu F_{\nu\mu}=0.
\end{equation}
If one takes the limit $q\rightarrow\infty$ keeping $q\rho_\mu$ and $q A_\mu$ fixed, the back reaction of the matter sources to the background can be ignored. This is the probe limit we will adopt in this paper.  The background is taken to be a $(3+1)$ dimensional Schwarzschild-AdS black hole
\begin{equation}\label{metric}
ds^2=-f(r)dt^2+\frac{dr^2}{f(r)}+r^2(dx^2+dy^2),
\end{equation}
with $f(r)=r^2(1-\frac{r_h^3}{r^3})$ and $r_h$ the horizon radius. The Hawking temperature for this black hole is $T=\frac{f'(r_h)}{4\pi}=\frac{3r_h}{4\pi}$, which sets the temperature of the boundary field theory. In the probe limit, the matter fields $\rho_\mu$ and $A_\mu$ can be treated as perturbations on the Schwarzschild-AdS black hole background.

\section{Adding a constant magnetic field}
\label{sect:ansatzB}
In Appendix~\ref{sect:superconducor}, we discussed the condensate of the vector field $\rho_{\mu}$ induced by the bulk electric field by turning on the scalar potential $A_t$ associated with the Maxwell field only. The normal phase corresponds to the black hole solution with a vanishing vector field $\rho_{\mu}$. As one lowers the temperature, the normal phase becomes unstable to developing non-trivial configuration of the vector field $\rho_{\mu}$. It gives non-zero vacuum expectation value of the dual vector operator, which breaks the U(1) gauge symmetry and the rotational symmetry in $x-y$ plane. The calculation of the optical conductivity reveals that there is a delta function at the origin for the real part of the conductivity, which means the condensed phase is indeed superconducting. For details, see Appendix~\ref{sect:superconducor}.

We now turn on a magnetic field to study how the applied magnetic field influences on the system. A consistent ansatz is as follows~\footnote{One can also consider the dyonic black hole background with gauge field fixed~\cite{Albash:2008eh}. $\rho_\mu$ is considered as a perturbation in such a background. In that case, the metric field $f(r)$ will depend on electric charge and magnetic field $B$.}
\begin{equation}\label{matterB}
\begin{split}
\rho_\nu dx^\nu=[\epsilon\rho_x(r,x)e^{ipy}+\mathcal{O}(\epsilon^3)]dx+[\epsilon\rho_y(r,x)e^{ipy}e^{i\theta}+\mathcal{O}(\epsilon^3)]dy,\\
A_\nu dx^\nu=[\phi(r)+\mathcal{O}(\epsilon^2)]dt+[Bx+\mathcal{O}(\epsilon^2)]dy,
\end{split}
\end{equation}
where $\rho_x(r,x)$, $\rho_y(r,x)$ and $\phi(r)$ are all real functions, $p$ is a real constant, the constant $\theta$  is the phase of $\rho_y$ and $B>0$ is the constant magnetic field
perpendicular to the $x-y$ plane. We have defined the deviation parameter $\epsilon$ from the critical point at which the condensate begins to appear.

The zeroth order of~\eqref{gauge} gives the equation of motion for $\phi$
\begin{equation}\label{eomphi}
\phi''(r)+\frac{2}{r}\phi'(r)=0.
\end{equation}
The asymptotic value of $\phi$ gives the chemical potential $\mu=A_t(r\rightarrow\infty)$ of the dual field theory.  The boundary condition at the horizon is given by requiring that $A_\mu A^\mu$ is finite there. Thus we can obtain a unique solution
\begin{equation}\label{phi}
\phi(r)=\mu(1-\frac{r_h}{r}).
\end{equation}

The equations of motion for $\rho_x$ and $\rho_y$ can be deduced from~\eqref{vector} at order $\mathcal{O}(\epsilon)$. We further separate the variables as $\rho_x(r,x)=\varphi_x(r)X(x)$ and $\rho_y(r,x)=\varphi_y(r)Y(x)$. We find that to satisfy the equations of motion of the model with the given ansatz, $\theta$ can only be chosen as $\theta_+=\frac{\pi}{2}+2n\pi$ or $\theta_-=-\frac{\pi}{2}+2n\pi$ with $n$ an arbitrary integer. The equations of motion for $\varphi_x(r)$, $\varphi_y(r)$, $X(x)$ and $Y(x)$ are divided into the following equations as
\begin{equation}\label{eomro1}
\varphi_x \dot{X}\pm (qBx-p)\varphi_y Y=0,
\end{equation}
\begin{equation}\label{eomro2}
\varphi_x' \dot{X}\pm (qBx-p)\varphi_y' Y=0,
\end{equation}
\begin{equation}\label{eomro3}
\varphi_x''+\frac{f'}{f}\varphi_x'+\frac{q^2\phi^2}{f^2}\varphi_x-\frac{m^2}{f}\varphi_x+\frac{\varphi_x}{r^2f}[\mp (qBx-p)\frac{\dot{Y}}{X}\frac{\varphi_y}{\varphi_x}\pm qB\gamma\frac{Y}{X}\frac{\varphi_y}{\varphi_x}-(qBx-p)^2]=0,
\end{equation}
\begin{equation}\label{eomro4}
\varphi_y''+\frac{f'}{f}\varphi_y'+\frac{q^2\phi^2}{f^2}\varphi_y-\frac{m^2}{f}\varphi_y+\frac{\varphi_y}{r^2f}[\frac{\ddot{Y}}{Y}\pm (qBx-p)\frac{\dot{X}}{Y}\frac{\varphi_x}{\varphi_y}
\pm (1+\gamma)qB\frac{X}{Y}\frac{\varphi_x}{\varphi_y}]=0,
\end{equation}
where the prime denotes the derivative with respect to $r$ and the dot denotes the derivative with respect to $x$. Here and below the upper signs correspond to the $\theta_+$ case and the lower to the $\theta_-$ case.
In order to satisfy~\eqref{eomro1}, one should impose
\begin{equation}\label{condition1}
\varphi_y=c\varphi_x, \   \ \dot{X}\pm c(qBx-p)Y=0,
\end{equation}
where $c$ is a real constant. This constraint is automatically satisfied by~\eqref{eomro2}.  We can see that only two of the four functions are independent. Substituting~\eqref{condition1} into the remaining equations, we can find the following three equations
\begin{equation}\label{eomroa}
\varphi_x''+\frac{f'}{f}\varphi_x'+\frac{q^2\phi^2}{f^2}\varphi_x-\frac{m^2}{f}\varphi_x-\frac{E}{r^2f}\varphi_x=0,
\end{equation}
\begin{equation}\label{eomrob}
-\ddot{X}\mp c(1+\gamma)qBY+(qBx-p)^2X=EX,
\end{equation}
\begin{equation}\label{eomroc}
-\ddot{Y}\mp \frac{(1+\gamma)qB}{c}X+(qBx-p)^2Y=EY.
\end{equation}

One can get the value of $E$ for arbitrary constant $c$ by solving the eigenvalue problem~\eqref{eomrob} and~\eqref{eomroc} with the constraint given in~\eqref{condition1}. There may exist the possibility that one can not obtain non-trivial solutions for some special values of $c$.  Here we consider a simple case with $c^2=1$, in which the equations of motion for $X(x)$ and $Y(x)$ can be solved exactly. Since $c=1/c$, the $c$ in the denominator in~\eqref{eomroc} is equivalent to the case in the numerator. Subtracting~\eqref{eomrob} from~\eqref{eomroc} and defining a new function as~\footnote{To solve the above equations, one can also define $\psi(x)=X(x)+Y(x)$. This case is equivalent to setting $c\rightarrow-c$, which gives nothing new.}
\begin{equation}\label{psi}
\psi(x)=X(x)-Y(x),
\end{equation}
one gets the equation
\begin{equation}\label{eompsi}
\ddot{\psi}(x)+[E\mp c(1+\gamma)qB-(qBx-p)^2]\psi(x)=0.
\end{equation}
We further introduce a new variable $\xi=\sqrt{|qB|}(x-\frac{p}{qB})$ and a constant $\eta=\frac{E\mp c(1+\gamma)qB}{|qB|}$, then the above equation becomes
\begin{equation}\label{eompsi1}
\frac{d^2}{d\xi^2}\psi(\xi)+(\eta-\xi^2)\psi(\xi)=0.
\end{equation}
The regular and bounded solution of~\eqref{eompsi1} is given by Hermite function $H_n$ as
\begin{equation}\label{solutionpsi1}
\psi_n(\xi)=N_n e^{-\xi^2/2}H_n(\xi),
\end{equation}
with the corresponding eigenvalue $\eta_n=2n+1$. $N_n$ is a normalization constant and $n$ is a non-negative integer.  Thus we obtain the solution of equation \eqref{eompsi} as
\begin{equation}\label{solutionpsi}
\psi_n(x)=N_n e^{-\frac{1}{2}|qB|(x-\frac{p}{qB})^2}H_n(\sqrt{|qB|}(x-\frac{p}{qB})),
\end{equation}
with the corresponding eigenvalue
\begin{equation}\label{eigenvalue}
E_n=(2n+1)|qB|\pm c(1+\gamma)qB.
\end{equation}

Combining~\eqref{condition1},~\eqref{psi} and~\eqref{solutionpsi}, one can obtain the exact configurations for $X(x)$ and $Y(x)$, which read
\begin{equation}\label{solutionx}
\begin{split}
X_n(x;p)=e^{\mp\frac{cqB}{2}(x-\frac{p}{qB})^2}&[X(0)e^{\pm\frac{cp^2}{2qB}}\pm cqBN_n\times\\
&\int_0^x (t-\frac{p}{qB})e^{-\frac{|qB|\mp cqB}{2}(t-\frac{p}{qB})^2}H_n(\sqrt{|qB|}(t-\frac{p}{qB}))dt],
\end{split}
\end{equation}
and
\begin{equation}\label{solutiony}
Y_n(x;p)=X_n(x;p)-N_n e^{-\frac{1}{2}|qB|(x-\frac{p}{qB})^2}H_n(\sqrt{|qB|}(x-\frac{p}{qB})),
\end{equation}
where $X(0)$ is a constant denoting the value of $X_n(x;p)$ at the origin $x=0$. The solutions of $\varphi_x$ and $\varphi_y$ corresponding to the eigenvalue $E_n$, denoted by $\varphi_{xn}$ and $\varphi_{yn}$, can be obtained by solving the equation of motion~\eqref{eomroa} with $E_n$ given in~\eqref{eigenvalue}. So far, we have recovered the Landau levels. As one can see in appendix~\ref{sect:superconducor}, we have a second order phase transition from the normal phase with  $\rho_\mu=0$ to the condensed phase with $\rho_{\mu}\ne 0$. Therefore we should encounter a marginally stable mode at the transition point. Theses solutions obtained from~\eqref{eomroa} just correspond to the marginally stable states.

We can see from~\eqref{eomroa} that the effective mass of $\rho_x$ is
\begin{equation}\label{mass}
m_{\rm eff}^2=m^2+\frac{E_n}{r^2}-\frac{q^2\phi^2}{f}=m^2+\frac{(2n+1)|qB|\pm c(1+\gamma)qB}{r^2}-\frac{q^2\phi^2}{f},
\end{equation}
which is clearly shifted by the magnetic field $B$. Depending on concrete parameters and Landau level, the appearance of magnetic field can increase or decrease the effective mass, thus will hinder or enhance the transition from the normal phase to the condensed phase. In what follows, we consider the case with the lowest Landau level with $n=0$, which reads
\begin{equation}\label{solution1}
\begin{split}
E_0^L=-|\gamma qB|,\\
X_0^L(x;p)=\frac{N_0}{2}e^{-\frac{|qB|}{2}(x-\frac{p}{qB})^2}=-Y_0^L(x;p).
\end{split}
\end{equation}
%
%\begin{equation}\label{solution2}
%\begin{split}
%E_b=2|qB|+|\gamma qB|,
%X_b(x)=e^{-\frac{|qB|}{2}(x-\frac{p}{qB})^2}[X(0)e^{\frac{p^2}{2|qB|}}-\frac{N_0 p^2}{2|qB|}+\frac{N_0|qB|}{2}(x-\frac{p}{qB})^2].
%\end{split}
%\end{equation}
%
%
\section{ Phase diagram}
\label{sect:diagram}
We are interested in how the applied magnetic field influences on the transition temperature from the normal phase to the condensed phase. As we can see from~\eqref{mass}, the effective mass of the charged vector field $\rho^\mu$ in the lowest energy state, i.e., in the lowest Landau level $n=0$ depends on the magnetic field $B$ and the non-minimal  coupling parameter $\gamma$ as
\begin{equation}\label{lowestmass}
m_{\rm eff}^{2}=m^2-\frac{|\gamma qB|}{r^2}-\frac{q^2\phi^2}{f}.
\end{equation}
It is clear that the increase of the magnetic field $B$ decreases the effective mass and thus tends to raise the transition temperature.
\subsection{Phase diagram at vanishing charge density}
It is well known that the increase of the electric field decreases the effective mass of charged scalar or vector fields, inducing the transition from the normal phase to the condensed phase. We first turn off the electric field, which corresponds to the case with vanishing charge density $\rho=0$. We introduce a new coordinate $z=r_h/r$. The equation of motion~\eqref{eomroa} can be rewritten as~\footnote{We apologize to the readers for here using a same notation to denote the derivative with respect to different variables for brevity. But the meaning of the derivative in the text is clear and will not be confused.}
\begin{equation}\label{eomroa1}
\varphi_x''(z)-\frac{3z^2}{1-z^3}\varphi_x'(z)-[\frac{m^2}{z^2(1-z^3)}-\frac{9\zeta}{16\pi^2(1-z^3)}]\varphi_x(z)=0,
\end{equation}
with $\zeta=|\gamma qB|/T^2$.

To solve such a second order equation by shooting method, we impose the regular condition at the horizon $z=1$ as well as the source free condition at the boundary $z=0$. More specifically, we set $\varphi_x(1)=1$ in our numerical calculation due to the linearity of~\eqref{eomroa1}. For a given $m^2$, only for certain values of $\zeta=|\gamma qB|/T^2$ can the boundary conditions be satisfied.
\begin{figure}[h]
\centering
\includegraphics[scale=0.90]{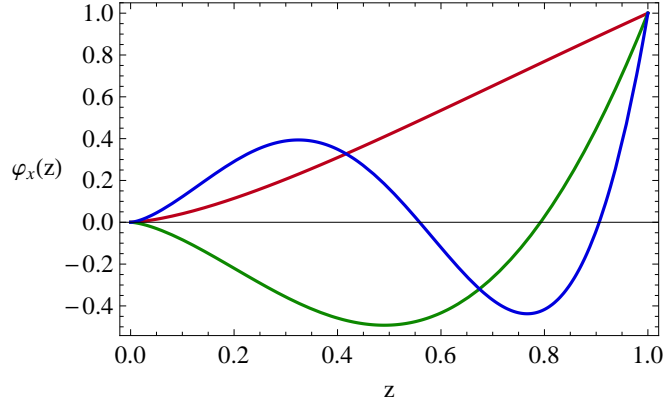}\caption{\label{eigenvalue1} The marginally stable curves of the charged vector field corresponding to various critical $\zeta=|\gamma qB|/T^2$. The three curves from top to down correspond to $\zeta_0\simeq72.84$ (red), $\zeta_1\simeq339.24$ (green) and $\zeta_2\simeq780.10$ (blue), respectively. We choose $m^2=3/4$.}
\end{figure}

\begin{figure}[h]
\centering
\includegraphics[scale=0.9]{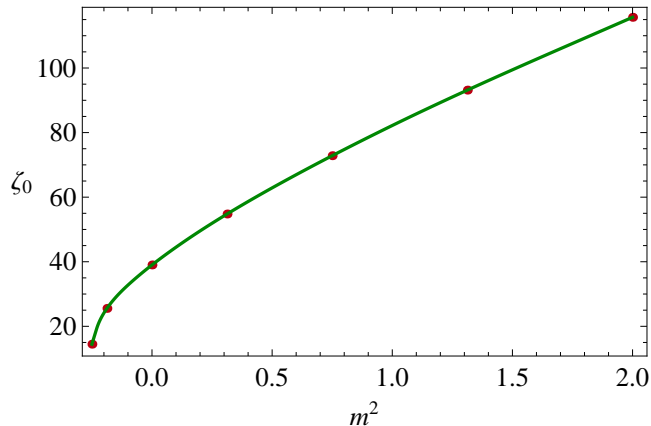}\caption{\label{eigenvaluem} The critical magnetic field $\zeta_0$  versus $m^2$ of the vector field. The points are obtained by the shooting method to solve the equation~\eqref{eomroa1}.}
\end{figure}

Figure~\ref{eigenvalue1} presents the three marginally stable curves of $\varphi_x(z)$ for $m^2=3/4$. The three lowest-lying modes are in the sequence $\zeta_0 < \zeta_1 < \zeta_2$. The red line corresponding to the minimal value of $\zeta$ has no intersecting points with horizontal axis at non-vanishing $z$. Such a mode with $\zeta_0\simeq72.84$ is considered as a mode of node $n=0$. Furthermore, the green line  to $\zeta_1\simeq339.24$ and blue line to $\zeta_2\simeq780.10$ are regarded as modes with nodes $n=1$ and $n=2$, respectively. Since the radial oscillations in $z$-direction of $\varphi_x(z)$ will cost more energy, the later two curves are therefore thought to be unstable.  Thus the lowest value $\zeta_0$ just gives the critical magnetic field above which the normal state is unstable to developing a vector hair. Figure~\ref{eigenvaluem} shows the critical magnetic field in terms of $\zeta_0$ for various squared mass of the vector field. It can be seen clearly that $\zeta_0$ increases as we increase the squared mass.

It should be pointed out that if we turn off the magnetic field, the normal state will not become unstable to developing hairs with vanishing charge density. The interesting result here is that only the magnetic field itself can trigger the phase transition. This result has an analogy to the QCD vacuum instability induced by a strong magnetic field to spontaneously developing the $\rho$-meson condensate. It is clear that the last term in~\eqref{action} describing a non-minimal coupling of the vector field $\rho^\mu$ to the gauge field $A_\mu$ plays a crucial role in  the instability. Note that similar coupling can be found in many formalisms used to describe the coupling of magnetic moment to the background magnetic field for charged particles of spin 1, i.e., vector particles~\cite{Djukanovic:2005ag, Young:1963zza}.

\subsection{Phase diagram at finite charge density}
Let us now consider the system with fixed charge density $\rho$. Here we do not limit ourselves to the lowest energy state. The equation of motion~\eqref{eomroa} can be written in terms of $\rho$ and $z$ as
\begin{equation}\label{eomroa2}
\varphi_x''(z)-\frac{3z^2}{1-z^3}\varphi_x'(z)-[\frac{m^2}{z^2(1-z^3)}+\frac{(E_n/\rho)\lambda}{1-z^3}-\frac{q^2\lambda^2}{(1+z+z^2)^2}]\varphi_x(z)=0,
\end{equation}
where $\lambda=\frac{\rho}{r_h^2}$ and $E_n$ is given in~\eqref{eigenvalue}. The lowest energy state corresponds to $E_n=E_0^L=-|\gamma qB|$. For numerical convenience and to match the behavior at the boundary, we further define
\begin{equation}\label{FF}
\varphi_x(z)=(\frac{z}{r_h})^{\triangle_-}F(z).
\end{equation}
Then we can obtain
\begin{equation}\label{eomF}
\begin{split}
F''(z)-\frac{1}{z}(\frac{3z^3}{1-z^3}-2\triangle_-)F'(z)-[\frac{m^2}{z^2(1-z^3)}+\frac{3{\triangle_-}z}{1-z^3}-\frac{{\triangle_-}({\triangle_-}-1)}{z^2}]F(z)\\
+[\frac{q^2\lambda^2}{(1+z+z^2)^2}-\frac{(E_n/\rho)\lambda}{1-z^3}]F(z)=0.
\end{split}
\end{equation}
\begin{figure}[h]
\centering
\includegraphics[scale=1]{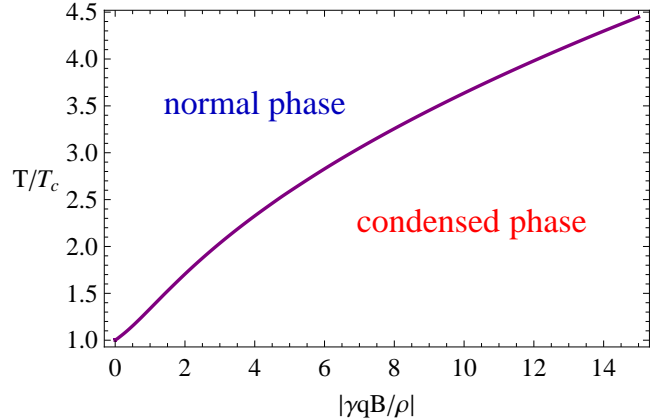}\caption{\label{magneticn}  The transition temperature from the normal phase to the condensed phase as a function of magnetic field. It corresponds to the case with  $E_n=-|\gamma qB|$. $T_c$ is the critical temperature in the case without magnetic field. The magnetic field raises the transition temperature. We choose $m^2=3/4$.}
\end{figure}

This equation depends on two physical parameters $\frac{T}{\sqrt{\rho}}=\frac{3}{4\pi \sqrt{\lambda}}$ and $\frac{E_n}{\rho}$. To solve such second order equation, we impose the regular condition at the horizon $z=1$ as well as the source free condition $F(0)=0$ at the boundary $z=0$. It has a non-trivial solution only when there is a relation between such two parameters, which just gives the transition temperature as a function of the magnetic field.

\begin{figure}[h]
\centering
\includegraphics[scale=1]{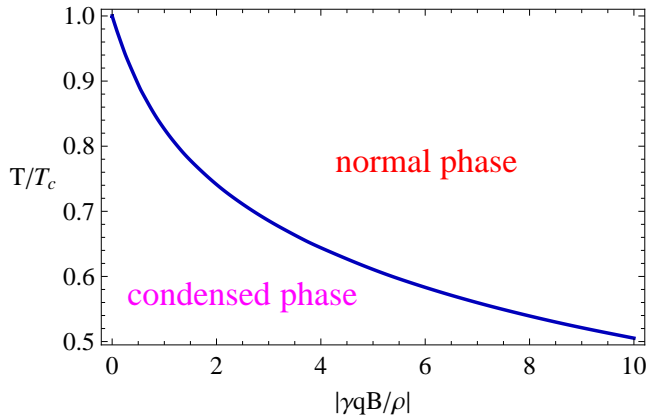}\caption{\label{magneticp}  The transition temperature from the normal phase to the condensed phase versus  magnetic field. It corresponds to the case with $E_n=|\gamma qB|$. $T_c$ is the critical temperature in the case without magnetic field. The magnetic field leads to lowering the transition temperature. We choose $m^2=3/4$.}
\end{figure}

The $(T, B)$ phase diagram with the lowest Landau level is drawn in figure~\ref{magneticn}. To determine which side of the phase transition line is the condensed phase, we can consider the equation~\eqref{lowestmass}. It suggests that the magnetic field decreases the effective mass. So if we increase the magnetic field at a fixed temperature, the normal state will become unstable for sufficiently large magnetic field. Figure~\ref{magneticn} looks very similar to figure~9 in ref.~\cite{Callebaut:2013ria} where the chiral transition temperature rises with magnetic field, indicating chiral magnetic catalysis. Furthermore, to compare with the lowest Landau level case, we present an example with positive $E_n=|\gamma qB|$ in~\eqref{eomroa2} in figure~\ref{magneticp}. One can see clearly in this case that the transition temperature lowers with the increase of the applied magnetic field. It is the well known property of the ordinary superconductor, which has been first discussed in a holographic setup in ref.~\cite{Maeda:2009vf}.

\section{Vortex lattice solution}
\label{sect:vortex}
Let us now construct the vortex lattice  solution. It is enough to consider  the $n=0$ solution only, i.e.,
\begin{equation}\label{zero}
\psi_0(x;p)=N_0e^{-\frac{1}{2}|qB|(x-\frac{p}{qB})^2}.
\end{equation}
Due to the fact that the eigenvalue $E_n$ is independent of $p$, a linear superposition of the solutions $e^{ipy}\varphi_{xn}(r)X_n(x;p)$ and $e^{ipy}\varphi_{yn}(r)Y_n(x;p)$ with different $p$ is also a solution of the model at $\mathcal{O}(\epsilon)$. We introduce two functions
\begin{equation}\label{newfunctions}
\begin{split}
\rho_x(r,x,y)=\varphi_{x0}(r)\sum_{\ell=-\infty}^{+\infty}c_\ell e^{ip_\ell y}X_0^L(x;p_\ell),\\
\rho_y(r,x,y)=ce^{i\theta_\pm}\varphi_{x0}(r)\sum_{\ell=-\infty}^{+\infty}c_\ell e^{ip_\ell y}Y_0^L(x;p_\ell),\\
c_\ell=e^{-i\frac{\pi a_2}{a_1^2}\ell^2},\ \ \ \ \ \ p_\ell=\frac{2\pi\sqrt{|qB|} \ell}{a_1},
\end{split}
\end{equation}
which satisfy the full equations of motion. $a_1$ and $a_2$ are arbitrary constants. Following  ref.~\cite{Maeda:2009vf}, we can obtain the vortex lattice solution
\begin{equation}\label{superposition}
\triangle\rho(r,x,y)\equiv\rho_x(r,x,y)-ce^{-i\theta_\pm}\rho_y(r,x,y)=\varphi_{x0}(r)\sum_{\ell=-\infty}^{+\infty}c_\ell e^{ip_\ell y}\psi_0(x;p_\ell).
\end{equation}
$\triangle\rho(r,x,y)$ has a pseudo-periodicity
\begin{equation}\label{property1}
\begin{split}
\triangle\rho(r,x,y)=\triangle\rho(r,x,y+\frac{a_1}{\sqrt{|qB|}}),\\
\triangle\rho(r,x+\frac{2\pi}{a_1\sqrt{|qB|}},y+\frac{a_2}{a_1\sqrt{|qB|}})=e^{\frac{i2\pi}{a_1}(\sqrt{|qB|}y+\frac{a_2}{2a_1})}\triangle\rho(r,x,y),
\end{split}
\end{equation}
as well as a zero at
\begin{equation}\label{property2}
\textbf{x}_{m,n}=(m+\frac{1}{2})\textbf{b}_1+(n+\frac{1}{2})\textbf{b}_2,
\end{equation}
with two vectors $\textbf{b}_1=\frac{a_1}{\sqrt{|qB|}}\partial_y$ and $\textbf{b}_2=\frac{2\pi}{a_1\sqrt{|qB|}}\partial_x+\frac{a_2}{a_1\sqrt{|qB|}}\partial_y$. $m$ and $n$ are two integers.

Since the expectation value of the operator $\hat{J^\mu}$ dual to $\rho_\mu$ is given by the coefficient at boundary $r\rightarrow\infty$, the quantity $J_{\pm}=\langle\hat{J^x}\pm i\hat{J^y}\rangle$ indeed exhibits the vortex structure with the cores of vortices located at $\textbf{x}_{m,n}$. In particular, the triangular lattice with three adjoining vortices forming an equilateral triangle can be obtained by choosing the following parameters
\begin{equation}\label{triangular}
a_1=\frac{2\sqrt{\pi}}{\sqrt[4]{3}},\    \ a_2=\frac{2\pi}{\sqrt{3}}.
\end{equation}
It should be stressed that it is the special combinations $J_{\pm}$ which exhibit the vortex lattice structure. In particular, for $q>0$, it is $J_-$ that corresponds to the lowest Landau level, while for $q<0$, it is  $J_+$. The form of operator presenting vortex lattice structure is the same as the one in field theory study without gravity~\cite{Chernodub:2010qx}. This also provides the evidence for the correctness of choosing $c^2=1$ in section~\ref{sect:ansatzB}.

\begin{figure}[h]
\centering
\includegraphics[scale=0.46]{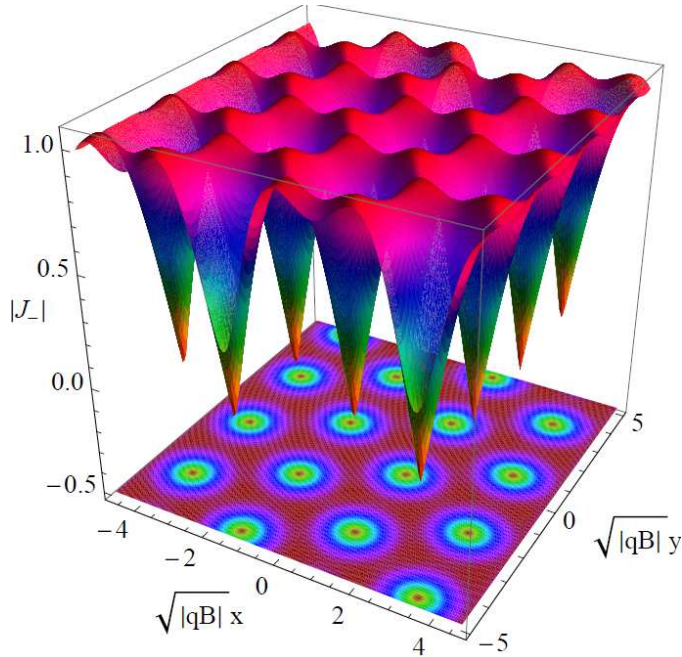}\caption{\label{lattice} The vortex lattice structure for the triangular lattice in $x-y$ plane. The contour plot is also drawn in the bottom. In particular, the condensate vanishes in the core of each vortex.}
\end{figure}

Figure~\ref{lattice} shows the configuration of the norm of condensate $J_{-}$ in the $x-y$ plane for the triangular lattice. Obviously, to obtain the true ground state, we should calculate the free energy of the solutions with different lattice structures from the action to find which configuration minimizes the free energy. It turns out that the linear analysis presented here is not sufficient to determine the most stable solution. We should include higher order contributions just as done in refs.~\cite{Bu:2012mq,Maeda:2009vf}. The calculation is much more complicated and is not very relevant to our purpose of this paper. We leave it for our further study.
%
%
%%%%%%%%%%%%%%%%%%%%%%%%%%%%%%%%%%%%Conclusion%%%%%%%%%%
%%%%%%%%%%%%%%%%%%%%%%%%%%%%%%%%%%%%%%%%%%%%%%%%%Conclusion%%%%%%%%%
%%%%Conclusion%%%%%%%%

\section{Conclusion and discussion}
\label{sect:conclusion}
In this paper we studied a holographic model with a complex vector field charged under a U(1) gauge field in a $(3+1)$ dimensional AdS black hole background, aiming to shed some light on the real strongly coupled systems which are of gravity duals. In this model, there is a non-minimal coupling of the vector field to the U(1) gauge field, which describes the interaction between the magnetic dipole moment of the vector field to the background magnetic field. This model includes the minimal ingredients to build a holographic
 p-wave superconductor model. We found a critical temperature at which the system undergoes a second order phase transition. The critical exponent of this transition is one half which coincides with the case in the Landau-Ginzburg theory. In the condensed phase, a vector operator acquires a vacuum expectation value breaking the Abelian symmetry as well as rotational symmetry spontaneously. Our calculation showed that this condensed phase exhibits an infinite DC conductivity and a gap in the frequency-dependent conductivity, which is quite similar to properties of the ordinary superconductivity.

We paid more attention on the response of this system to an applied magnetic field. We obtained the Landau level, from which we can find the contribution to the effective mass of the vector field by the magnetic field (see~\eqref{mass}). Due to the non-minimal coupling given in the last term of~\eqref{action}, the applied magnetic field can reduce the effective mass of the vector field, thus inducing the instability of the black hole background even when the chemical potential/charged density is absent. That is, for the case with vanishing chemical potential or charged density, the black hole background becomes no longer stable when the magnetic field is beyond a certain critical value. For the case with non-vanishing chemical potential, the phase boundary is determined by a relation between the transition temperature and the magnetic field, which is presented in figure~\ref{magneticn}. The transition temperature increases with the applied magnetic field. The response of this system to the magnetic field is quite different from the behavior of ordinary superconductor where the magnetic field makes the transition more difficult as drawn in figure~\ref{magneticp}.  But our result is quite similar to the case of QCD vacuum instability induced by strong magnetic field to spontaneously developing the $\rho$-meson condensate~\cite{Chernodub:2010qx,Chernodub:2011mc}. Although so, it was shown that in our model, the condensate of the vector operator forms a vortex lattice structure in the spatial directions perpendicular to the magnetic field. Of course, the non-minimal coupling term in the action plays a crucial role in both cases. Therefore in some sense, our model is a holographic setup of the study of $\rho$-meson condensate in refs.\cite{Chernodub:2010qx,Chernodub:2011mc}.

In ordinary superconductors an external magnetic field suppresses superconductivity via diamagnetic and Pauli pair breaking effects. However, it has also been proposed that the magnetic field induced superconductivity can also be realized in type-II superconductors~\cite{PhysRevLett.58.1482,Rasolt:1992zz}, in which the Abrikosov flux lattice may enter a quantum limit of the low Landau level dominance with a spin-triplet pairing. And possible experimental evidence for the strong magnetic induced superconductivity can be found, for example, in refs.~\cite{levy2005,uji2010}. It was also shown in Gross-Neveu type model that applied magnetic field might induce superconductivity in the planar system with 4-fermion interaction~\cite{Klimenko:2012qi}.

We mention here that similar studies can also be generalized other gravitational backgrounds, such as the AdS soliton background which has been adopted to mimic superconductor/insulator phase transition~\cite{Nishioka:2009zj}. The study of magnetic field effect in the superconductor/insulator case can be found in ref.~\cite{Cai:2011tm}. In a forthcoming paper~\cite{Cai:2013kaa}, we generalize the present study to the case with the AdS soliton as the background. It is found that the magnetic field can  induce the instabilities of the AdS soliton background. Comparing our model with the SU(2) model with a constant non-Abelian magnetic field~\cite{Ammon:2011je,Wong:2013rda}, we find that our complex vector field model in some sense is a generalization of the SU(2) model to the case with a general mass squared $m^2$ and magnetic moment characterized by $\gamma$. In the setup of the present paper, the SU(2) model corresponds to our model with $m^2=0$ and $\gamma=1$.

In this paper, we restricted ourselves to the probe approximation, neglecting the effect of matter fields on the background geometry. This can indeed reveal some significant properties of the model, but something might be lost in this approximation, see ref.~\cite{Cai:2013wma} as an example. It is therefore helpful to understand full properties of the model by considering the back reaction of matter fields on the background geometry. In a recent paper~\cite{Cai:2013aca}, going beyond the probe approximation, we found a rich phase structure in this model without magnetic field. Depending on mass square $m^2$ and charge $q$ of the vector field,  not only second order but also  first order and zeroth order phase transition can appear. Interestingly, there also exists a so-called ``retrograde condensation'' in which the hairy solution exists only for temperatures above a critical value and is thermodynamically subdominant. Particularly, the zeroth order transition and retrograde condensation can be observed for the case with small $m^2$. Indeed this model has much more phase behaviors than the SU(2) model, thus it can be used to mimic much richer phenomena in dual strongly coupled systems.

In our model, it is the magnetic field itself that can induce the condensate. A natural question arises how about the Meissner effect known as that superconductors expel weak external magnetic field. This effect is due to the large superconducting currents induced in the superconductor by the external magnetic field, which generates a back-reacting magnetic field screening the external magnetic field. In the holographic SU(2) p-wave model, it has been pointed out~\cite{Wong:2013rda} that the vortex currents flow in the opposite direction to the one in conventional superconductors, thus can enhance the applied magnetic field in the regions between the vortices. It seems that the model presented here is similar to the SU(2) p-wave model. It is interesting to ask whether a similar phenomenon also appears in our model.  To answer this question, we need to consider the contribution of higher order terms in~\eqref{matterB}, which is also required to find the true vortex flux lattice. We leave it for further study.

 In this paper we only calculated the conductivity in the case without magnetic field in one spatial direction. Although it is enough to see the superconductivity feature of the condensed phase, in order to study the model in a more realistic manner, it is desirable to calculate the conductivity in another direction and further to study the transport properties of the lattice state.  As a phenomenological approach, this toy model would have potential application to mimic strongly coupled systems with a vector like order parameter. It should be applicable in a wide variety of condensed matter systems, heavy ion physics and beyond. We hope to report further progresses in future.

\section*{Acknowledgements}

This work was supported in part by the National Natural Science Foundation of China (No.10821504, No.11035008, No.11205226,No.11305235, and No.11375247), and in part by the Ministry of Science and Technology of China under Grant No.2010CB833004.
\appendix

\section{Condensate of vector field and superconducting phase transition}
\label{sect:superconducor}
As a toy model with a charged vector field in the bulk dual to a vector operator in the field theory, there exists the possibility that the condensate of this vector field can serve as an order parameter to mimic a holographic p-wave superconductor phase transition, like the s-wave case~\cite{Hartnoll:2008vx,Hartnoll:2008kx}. More precisely, we hope that this system has stable black hole solutions with vector hair at low temperatures, but without vector hair at high temperatures. If this is true, in the condensed phase, the condensate of the dual vector operator will break not only the U(1) symmetry but also the rotational symmetry since the condensate of vector field  picks out one special direction. This situation is very similar to the one in the holographic p-wave superconductor model with SU(2) gauge field in ref.~\cite{Gubser:2008wv}. In this sense, our model can also be regarded as a holographic p-wave model.
\subsection{Condensate of vector field}
\label{sect:phase}
We adopt the following ansatz
\begin{equation}\label{matter}
\rho_\nu dx^\nu=\rho_x(r)dx+\rho_y(r)dy,\quad A_\nu dx^\nu=\phi(r)dt.
\end{equation}
One can use the U(1) gauge symmetry to set $\rho_x$ to be real. Then one finds that the $r$ component of~\eqref{gauge} implies that the phase of $\rho_y$ must be constant. Without loss of generality, we take $\rho_y$ to be real. Then, the independent equations of motion in terms of the above ansatz are deduced as follows
\begin{equation}\label{eoms}
\begin{split}
\rho_x''+\frac{f'}{f}\rho_x'+\frac{q^2\phi^2\rho_x}{f^2}-\frac{m^2\rho_x}{f}=0, \\
\rho_y''+\frac{f'}{f}\rho_y'+\frac{q^2\phi^2\rho_y}{f^2}-\frac{m^2\rho_y}{f}=0,\\
\phi''+\frac{2}{r}\phi'-\frac{2q^2}{r^2f}(\rho_x^2+\rho_y^2)\phi=0,\\
\end{split}
\end{equation}
where the prime denotes the derivative with respect to $r$.

In order to find the solutions for all the three functions $\mathcal{F}=\{\rho_x,\rho_y,\phi\}$ one must impose suitable boundary conditions at the AdS boundary $r\rightarrow\infty$ and at the horizon $r=r_h$. In addition to $f(r_h)=0$, one must require $\phi(r_h)=0$ in order for $g^{\mu\nu}A_\mu A_\nu$ to be finite at the horizon.

In order to match the asymptotical AdS boundary, the general falloff of the matter fields near the boundary $r\rightarrow\infty$ should behave as
\begin{equation} \label{boundary}
\begin{split}
\phi&=\mu-\frac{\rho}{r}+\ldots,\quad \rho_x=\frac{{\rho_x}_-}{r^{{\Delta}_-}}+\frac{{\rho_x}_{+}}{r^{{\Delta}_+}}+\ldots, \quad \rho_y=\frac{{\rho_y}_-}{r^{{\Delta}_-}}+\frac{{\rho_y}_{+}}{r^{{\Delta}_+}}+\ldots,
\end{split}
\end{equation}
where ${\Delta}_\pm=\frac{1\pm\sqrt{1+4 m^2}}{2}$.~\footnote{The $m^2$ has a lower bound as $m^2=-1/4$ with ${\Delta}_+={\Delta}_-=1/2$, in which case there is a logarithmic term in the asymptotical expansion. Such a term should be considered as the source set to be vanishing~\cite{Horowitz:2008bn}.} We impose ${\rho_x}_-=0$ and ${\rho_y}_-=0$, since we want
the U(1) symmetry to be broken spontaneously. According to the AdS/CFT dictionary, up to a normalization, the coefficients $\mu$, $\rho$, ${\rho_x}_{+}$ and ${\rho_y}_{+}$ are interpreted as chemical potential, charge density and the $x$ and $y$ components of the vacuum expectation value of the vector operator $\hat{J^\mu}$ in the dual field theory, respectively.

There is a useful scaling symmetry in the equations of motion
\begin{equation} \label{scaling}
r\rightarrow\lambda r,\quad (t,x,y)\rightarrow{\lambda^{-1}}(t,x,y),\quad(\phi,\rho_x,\rho_y)\rightarrow\lambda(\phi,\rho_x,\rho_y),
\end{equation}
where $\lambda$ is an arbitrary positive constant. Under this symmetry, the revelent quantities transform as
\begin{equation} \label{transform}
T\rightarrow\lambda T,\quad \mu \rightarrow\lambda\mu,\quad \rho\rightarrow\lambda^2\rho, \quad ({\rho_x}_{+},{\rho_y}_{+})\rightarrow\lambda^{\Delta_++1}({\rho_x}_{+},{\rho_y}_{+}).
\end{equation}
\begin{figure}[h]
\centering
\includegraphics[scale=1]{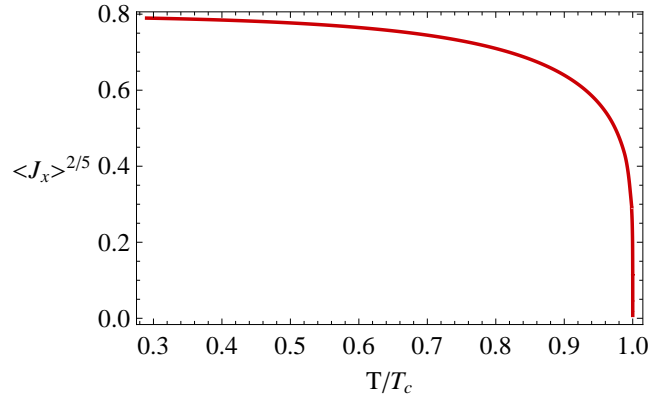}\caption{\label{condensate} The condensate as a function of temperature. We choose $q=1$ and $m^2=3/4$. The condensate begins to appear at $T_c\simeq0.102\sqrt{\rho}$.}
\end{figure}

We assume the condensate to pick out the $x$ direction as special, so we can consistently set $\rho_y=0$. The condensate as a function of temperature is presented in figure~\ref{condensate}. It is clear that as we lower the temperature, the normal phase with vanishing charged vector  becomes unstable to developing vector hair which breaks the U(1) symmetry as well as rotational symmetry in the dual field theory. Fitting the curve near the critical temperature $T_c$, we find that for small condensate there is a critical behavior with critical exponent $1/2$, which precisely meets  the result given by  mean field theory and it is typically a second order phase transition. In the case with $m^2=3/4$ we obtain
\begin{equation}
\langle\hat{J_x}\rangle\simeq339T_c^{5/2}(1-T/T_c)^{1/2},\   \ {\rm as}\   \ T\rightarrow T_c.
\end{equation}

Thus we have obtained two black hole solutions in the system. When $T>T_c$, we have the black hole solution without the vector field, while we have the black hole solution with
non-trivial vector field as $T<T_c$. This behavior is in complete agreement with the holographic superconducting phase transition in the literature. Therefore we expect that the black hole solution with non-trivial vector field can describe a superconducting phase. To prove this, it is helpful to calculate the optical conductivity. Before dong this, we should first ensure that the black hole solution with non-trivial vector field is more thermodynamically stable than the one without the vector field as $T<T_c$.
\subsection{Free energy}
\label{sect:free}
In order to determine which solution is thermodynamically favored, we should calculate the free energy of the system for both black hole solutions. We will work in canonical ensemble in this paper, where the charge density is fixed. In gauge/gravity duality Helmholtz free energy $F$ of the boundary thermal state is identified with temperature times the on-shell bulk action with Euclidean signature. Since we work in the probe approximation, we can ignore the gravity part. Given that the system is stationary, the Euclidean action is related to the Minkowski case by a total minus. Employing the equations of motion~\eqref{gauge} and~\eqref{vector}, we have
\begin{equation}\label{onshell}
\begin{split}
-2\kappa^2 S_{Euclidean}=\int d^4x\sqrt{-g}(-\frac{1}{4}F_{\mu\nu} F^{\mu \nu}-\frac{1}{2}\rho_{\mu\nu}^\dagger\rho^{\mu\nu}-m^2\rho_\mu^\dagger\rho^\mu+iq\gamma \rho_\mu\rho_\nu^\dagger F^{\mu\nu})\\+\int d^3x \sqrt{-h}n_\mu A_\nu F^{\mu\nu}+S_{ct}\\
=\int d^4x\sqrt{-g}\frac{1}{2}A_\nu\nabla_\mu F^{\mu \nu}+\int d^3x\sqrt{-h}n_\mu(\frac{1}{2}A_\nu F^{\mu\nu}-\rho_\nu^\dagger\rho^{\mu\nu})+S_{ct},
\end{split}
\end{equation}
where $h$ is the determinant of the induced metric $h_{\mu\nu}$ on the boundary $r\rightarrow\infty$ and $n^\mu$ is the outward pointing unit normal vector to the boundary. $S_{ct}$ denotes the surface counter term for removing divergence.

Substituting the asymptotically expansion~\eqref{boundary} into~\eqref{onshell} and introducing a counter term $S_{ct}=-\Delta_-\int dx^3\sqrt{-h}h_{\mu\nu}{\rho^\mu}^\dagger\rho^\nu$,~\footnote{We are not sure whether the counter term works or not in a general case. But we find this counter term works well for the ansatz~\eqref{matter} with the asymptotically expansion~\eqref{boundary}.} we find the free energy $F$ as
\begin{equation}\label{free}
\frac{2\kappa^2 F}{V}=\frac{1}{2}\mu\rho-(2\Delta_+-1)(\rho_{x-}\rho_{x+}+\rho_{y-}\rho_{y+})-\int_{r_h}^{\infty}dr\sqrt{-g}\frac{1}{2}A_\nu\nabla_\mu F^{\mu \nu},
\end{equation}
with $V=\int dxdy$.  Regarding $\rho_{x-}$ and $\rho_{y-}$ as sources, Helmholtz free energy tells us that sub-leading terms $\rho_{x+}$ and $\rho_{y+}$ are the expectation values of the dual operator in the field theory side.
\begin{figure}[h]
\centering
\includegraphics[scale=1.15]{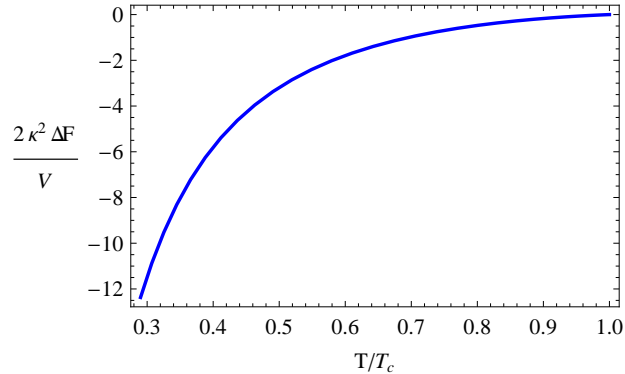}\caption{\label{freeen}  The difference of Helmholtz free energy between in the condensed phase  and in the normal phase as a function of temperature. We choose $q=1$ and $m^2=3/4$ for which the critical temperature $T_c\simeq0.102\sqrt{\rho}$.}
\end{figure}
The difference of Helmholtz free energy between the condensed phase and the normal phase as a function of temperature is presented in figure~\ref{freeen}. It is clear that below the critical temperature $T_c$, the state with non-vanishing vector ``hair" is indeed thermodynamically favored, compared to the normal phase. The phase transition is second order, which can be seen, for example, from the derivative of the free energy with respect to the temperature.
\subsection{Conductivity}
\label{sect:conduc}
We now calculate the conductivity in the dual field theory side as a function of frequency $\omega$. We need to turn on fluctuations of the matter contents in the bulk. We assume perturbations have a time dependence of the form $e^{-i\omega t}$ with zero spatial momentum. It turns out that one can consistently turn on the perturbations of $A_y$ only. We can obtain the equation of motion for $A_y$ by linearizing the equations of motion~\eqref{gauge}, which reads
\begin{equation}
A_y''+\frac{f^{\prime}}{f}A_y^{\prime}+(\frac{\omega^2}{f^2}-\frac{2 q^2\rho_x^2}{r^2 f})A_y=0.
\end{equation}
Since the conductivity is related to the retarded two-point function of the U(1) current, we impose the ingoing boundary condition near the horizon. The gauge field $A_y$ near the boundary $r\rightarrow\infty$ falls off as
\begin{equation}\label{axbound}
A_y=A^{(0)}+\frac{A^{(1)}}{r}+\cdots.
\end{equation}
According to the AdS/CFT dictionary, one can obtain the conductivity as
\begin{equation}\label{conduc}
\sigma_{yy}(\omega)=\frac{A^{(1)}}{i\omega A^{(0)}}.
\end{equation}
\begin{figure}[h]
\centering
\includegraphics[scale=0.81]{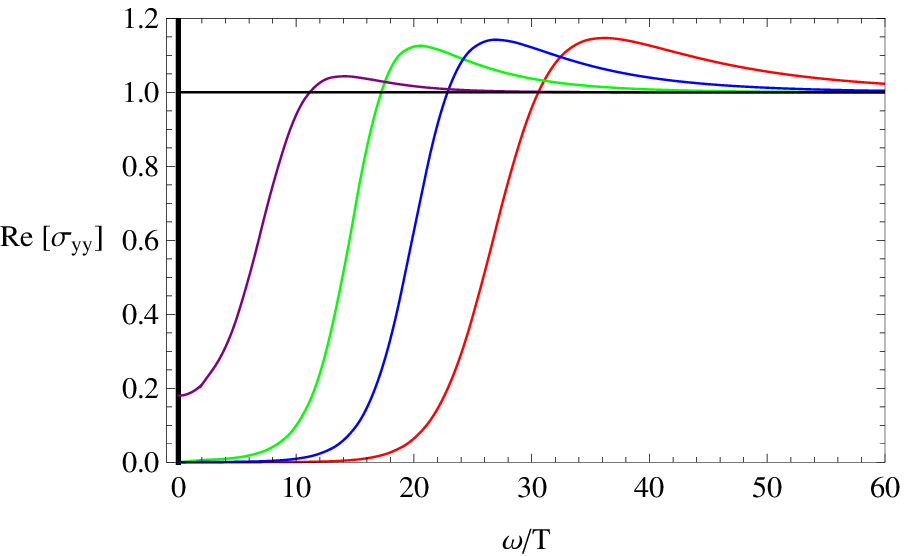}\ \ \ \
\includegraphics[scale=0.85]{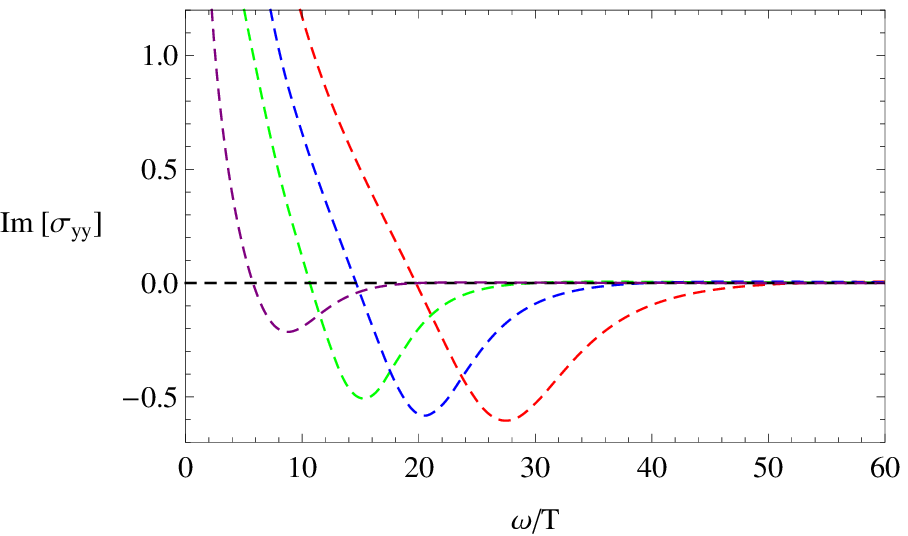} \caption{\label{conductivity} The optical conductivity as a function of frequence. The solid lines in the left plot are the real part of the conductivity, while the  dashed lines in the right plot are the imaginary part of the conductivity. We choose $q=1$ and $m^2=3/4$. The horizontal lines correspond to the temperature above $T_c$. Other curves from left to right correspond to $T/T_c\simeq0.830$ (purple), $T/T_c\simeq0.519$ (green), $T/T_c\simeq0.388$ (blue), and $T/T_c\simeq0.290$ (red), respectively. There is a delta function at the origin for the real part of the conductivity in the condensed phase.}
\end{figure}

The AC conductivity as a function of frequency is presented in figure~\ref{conductivity}.  We can see clearly that the optical conductivity along the $y$ direction in this model behaves qualitatively similar to the case in the p-wave model with SU(2) gauge symmetry~\cite{Gubser:2008wv}. In particular, from the Kramers-Kronig relation, one can conclude that the real part of the conductivity has a Dirac delta function at $\omega=0$ since the imaginary part has a pole, i.e., Im$[\sigma_{yy}(\omega)]\sim\frac{1}{\omega}$. Furthermore, it is clear that the optical conductivity develops a gap at some special frequency $\omega_g$ known as gap frequency. As suggested in ref.~\cite{Horowitz:2008bn}, it can be identified with the one at the minimum of the imaginary part of the AC conductivity. Re$[\sigma_{yy}]$ is very small in the infrared and rises quickly at $\omega_g$. There also exists a small ``bump" slightly above $\omega_g$ which is reminiscent of the behavior due to fermionic pairing~\cite{Gubser:2008wv}. For our chosen parameter, we have $\omega_g\simeq8T_c$. Compared to the corresponding BCS value $\omega_g\simeq3.5T_c$, the result shown here is consistent with the fact that our holographic model describes a system at strong coupling.

\end{document}